\documentstyle[12pt,epsfig,float,a4]{article}
\def\_{\rule{.3em}{.15ex}} 
\newcommand{\tb}  {\mbox{$ \tan\beta~ $}}

\newcommand{\be}{\begin{equation}}
\newcommand{\ee}{\end{equation}}
\newcommand{\bea}{\begin{eqnarray}}
\newcommand{\eea}{\end{eqnarray}}

\newcommand{\beq}{\begin{equation}}
\newcommand{\eeq}{\end{equation}}
\newcommand{\besg}{$b  \to  X_s \gamma~ $}

\def\slash#1{\setbox0=\hbox{$#1$}#1\hskip-\wd0\dimen0=5pt\advance
       \dimen0 by-\ht0\advance\dimen0 by\dp0\lower0.5\dimen0\hbox
         to\wd0{\hss\sl/\/\hss}}
\def\gequiv{\raise 0.4ex \hbox{$>$} \kern -0.7 em \lower 0.62 ex \hbox{$\sim$}}
\def\gappeq{\mathrel{\rlap {\raise.5ex\hbox{$>$}}
{\lower.5ex\hbox{$\sim$}}}}

\begin{document}

\begin{titlepage}

\begin{flushright}
IEKP-KA/2000-23 \\[3mm]
{\tt hep-ph/0007078}
\end{flushright}

\vspace*{-1cm}

\begin{center}
  {\large\bf The   \besg decay rate in NLO, \\
          Higgs boson limits, and LSP masses in the Constrained
Minimal Supersymmetric Model} \\[3mm]

  {\bf W. de Boer, M. Huber}
% H.-J. Grimm,N. Seibert, A. Sopczak}
\\[2mm]
  {\it Institut f\"ur Experimentelle Kernphysik, University of Karlsruhe \\
       Postfach 6980, D-76128 Karlsruhe, Germany} \\[3mm]

  {\bf A.V. Gladyshev, D.I. Kazakov} \\[2mm]

{\it Bogoliubov Laboratory of Theoretical Physics,
Joint Institute for Nuclear Research, \\
141 980 Dubna, Moscow Region, Russian Federation}

\end{center}

%\vspace{2cm}
\abstract{
New NLO \besg calculations
 have become available.
We observe that at large $\tb$ the dominant
NLO term of the chargino amplitude, which is
proportional to $\mu\tan^2\beta$, changes the sign of this
amplitude in a large
region of the CMSSM parameter space, so that the
preferred sign of the Higgs mixing parameter $\mu$ now agrees
with the preferred sign of $b-\tau$ unification.
We find that the \besg rate
does not constrain the CMSSM
anymore,
if    the  higher order contributions and its
uncertainties from the incomplete calculations  are  taken into account.
%Therefore  the high $\tb$ scenario  has
%in our global analysis
%a good $\chi^2$ again, which
%is  fortunate, since 
%the low $\tb$ CMSSM scenario with $\tb<3.3$   is excluded
%by the  Higgs mass limit above 107.9 GeV from LEP.h

The Higgs boson mass in the CMSSM is found to be
between 110 and 120 GeV for a top mass of 175 GeV, if the Higgs
mass limit of 107.9 GeV from LEP, which implies $\tb>3.3$,
 is taken into account 
The mean Higgs boson mass value and its dominant errors
are:
$m_h=115\pm3~ (stop mass)~\pm1.5~(stop mixing)~\pm2~(theory)~\pm5~(top
mass)~\rm GeV.$
This Higgs  mass range is valid for all
$\tb$ values above 20 and decreases  for lower $\tb$.

If the presently claimed evidence for dark matter
by the DAMA Collaboration is interpreted as
the  Lightest Supersymmetric Particle (LSP) of the CMSSM, then
it is  at the edge of the parameter space allowed by the present
Higgs limit of 107.9 GeV from LEP.

}

\end{titlepage}

\section{Introduction}

In a previous paper we showed that the inclusive decay rate \besg \  
severely constrains the high $\tan\beta$ solution of the
Constrained Minimal 
Supersymmetric Standard Model (CMSSM)~\cite{PL}.
This was mainly caused by 
the fact that $b-\tau$ Yukawa coupling unification
preferred a negative sign 
for the Higgs mixing parameter $\mu$,
while the \besg \  rate required 
the opposite sign. However, with the advent of 
next-to-leading order (NLO)
calculations for the \besg \  rate in the MSSM \cite{ciu} it
turns out that large terms proportional to $\tan^2\beta$
can change the sign of $\mu$.
Consequently the allowed parameter space becomes much larger for
the high $\tan\beta$ scenario, especially if the uncertainties
from the incomplete NLO calculations are taken into account in
our global analysis.

Here we used  the \besg \
rate from the
CLEO Collaboration, as presented at the
Stanford Lepton-Photon '99 Conference ~\cite{stanf}:
$
BR(b \to X_s \gamma) = ( 3.15 \pm 0.35 \pm 0.32 \pm 0.26 ) \cdot 10^{-4}.
$
This value 
combined with the less precise ALEPH measurement~\cite{ALEPH} of 
$
BR(b \to X_s \gamma) = (3.11 \pm 0.80 \pm 0.72 ) \cdot 10^{-4}
$
yields as average
$ BR(b \to X_s \gamma) = ( 3.14 \pm 0.48 ) \cdot 10^{-4}
$
%, since \besg
%does not constrain it more than the requirements of Electroweak
%Symmetry Breaking and gauge- and Yukawa coupling unification.
%In addition, it is shown that the theoretical uncertainties
%from the missing higher orders prevent any serious constraints on the CMSSM.

The present SM Higgs limit of 107.9 GeV\cite{newhiggs}
puts severe  constraints on the CMSSM parameter space, since
in the CMSSM the heavier Higgs decouple, so the lightest
Higgs has the properties of a SM Higgs.
These constraints, as well as the chargino limits from
LEP\cite{chargino}, are 
 compared with the recently claimed evidence by the DAMA
Collaboration\cite{dama} for dark matter, which can
be interpreted in the CMSSM as a stable neutralino with a
mass of $52^{+10}_{-8}$ GeV.
\section{Numerical analysis method}
Our statistical analysis of the allowed parameter space in
the MSSM\cite{PL} was repeated including the new partial
NLO \besg calculations.
In this $\chi^2$ analysis the constraints from gauge
coupling unification, $b-\tau$ Yukawa coupling unification,
electroweak symmetry breaking,  \besg, relic density  and
experimental lower limits on SUSY masses can be considered
either separately or combined.
In this paper we do not
include constraints from relic density, which are relevant mainly
at low $\tb$.

As free parameters of the Constrained MSSM (CMSSM) we consider
the unified gauge coupling constant ($\alpha_{\rm GUT}$) at the
unification scale (M$_{\rm GUT}$). In addition we define at the
GUT scale: 
%\begin{itemize}
%\item
the Yukawa coupling constants  of the third generation
(Y$_t^0$,Y$_b^0$,Y$_{\tau}^0$),
%\item
the common scalar mass (m$_0$),
%\item
 the common gaugino mass (m$_{1/2}$),
%\item
the common trilinear coupling (A$_t^0$=A$_b^0$=A$_{\tau}^0$),
%\item
 the ratio of the vacuum expectation values of the two Higgs doublets (\tb),
%\item
the Higgs mixing parameter $\mu^0$.
%\end{itemize}
The GUT scale parameters are optimized
  via a $\chi^2$ test to fit the low energy
experimental data on electroweak boson masses, \besg,
and quark and lepton masses of the third generation.

 The values of m$_0$, m$_{1/2}$, $\mu^0$, A$^0$,  Y$^0$ and $\tan\beta$
 determine completely the mass spectrum of all SUSY particles
 via the RGE. The values of $\mu^0$, Y$^0$
 and $\tan\beta$ are constrained for given values of m$_0$ and
 m$_{1/2}$ by EWSB and the quark and lepton masses of the third generation.
 Since m$_0$ and m$_{1/2}$ are strongly correlated, we
 repeat each fit for every pair of m$_0$ and
 m$_{1/2}$ values between (200,200) and (1000,1000) GeV in steps of 100 GeV.

As can be seen from Fig.~\ref{tbmtop} the value of \tb is constrained
by the present experimental value of the top mass
m$_t = 173.9\pm 5.2$ GeV\cite{pdb} to be in the following
ranges: 1$ < \tan\beta < $2 \  or
30$ < \tan\beta < $40
for $\mu < 0$.
These constraints result mainly from the $b-\tau$ Yukawa unification.
In the following we will first concentrate on these $\tb$ values,
which we call the low and high $\tb$ scenario,
but consider the complete $\tb$ dependence as well.
% The trilinear couplings at low energies tend to be independent
% of their values at the GUT scale due to their infrared fixed
% point (IRFP) behaviour.
% The GUT scale values of the trilinear couplings were found  to be
% insensitive to the \besg rate
%and the Higgs predictions
%and were set to zero. Note that the low energy values
%  of A$_t$,A$_b$ and A$_{\tau}$ are all different
% and nonzero and thus have to be taken into account, but they are not
% sensitive to the GUT starting values due to their infrared fixed point
% behaviour.

\section{ NLO  corrections to  \besg \ } 
The \besg transition corresponds in lowest order
to a loop with either a W, charged Higgs or chargino, 
as shown schematically  in Fig. \ref{f1}.
The leading order  corresponds to the emission of a
real photon from any of the charged lines, while the dominant
next-to-leading order (NLO) corrections involve virtual gluons
from any of the (s)quark lines,

The LO Standard Model (SM)
calculations~\cite{grin,BBMR,ajbur} have been complemented
by NLO calculations ~\cite{kadel,thurt1,aali,thurt2}.
Recently, the NLO calculations have been extended to
Two-Higgs Doublet Models (2HDM)~\cite{cdgg,borzu} and
the Minimal Supersymmetric Model (MSSM)
for a given mass hierarchy~\cite{ciu,mis}.
%However, it was assumed that only the lightest stop
%(predominantly righthanded) effectively contributes,
%while the other stop is too heavy. 
%In the CMSSM this is a bad approximation, since both stops
%have usually similar masses.
%%and were estimated to have
%%similar contributions.
%Including both masses can be done in a straightforward manner,
%if diagrams where both stops contribute, are neglected.
%%Since in the CMSSM the second stop has the same sign in NLO,
%Such calculations were done and lead to the same conclusion,
%i.e. a change of the preferred
%sign of $\mu$. However,  
%NLO calculations without approximations would be highly desirable
%to confirm this.

Here we use the results from Ref. \cite{ciu}, which are valid for
any value of $\tan\beta$.
After studying the paper from Ref. \cite{ciu} in detail\cite{huber},
we found that 
the NLO corrections to the chargino amplitude have 
two main contributions:
  the first one is proportional to $\mu\tb/\cos\beta$ 
(for large $\tb \propto \tan^2\beta$)
  coming from the sbottom mixing
 (e.g. the last diagram in Fig.\ref{f1}),
and the second one is proportional to large log terms
$\sim \log  {\tilde{m}_g}/{\mu_W}$
  coming from diagrams with gluinos.
(lowest  row in Fig.\ref{f1}).
The $\tan^2\beta$ dependence of the first term implies that these corrections
are significant for the high $\tb$ scenario discussed above. Note that the NNLO contributions do not have a $\tan^3\beta$, since
the sbottom mixing comes in only once, so the series is converging
rapidly afterwards, because of the $(\alpha_s/\pi)^2$ suppression.

The large NLO contributions
    change  the sign of the
chargino amplitude at large $\tb$ for practically the whole
parameter space, as shown in Fig. \ref{f3}.
%The strong $\tb$ dependence is displayed in Fig. \ref{f2}.
  Since the chargino-stop amplitude is of the same order of magnitude as
  the SM W-t amplitude for most of the parameter space,
    it interferes strongly: positively for $\mu>0$ and negatively
  for $\mu<0$. In the first case the \besg rate becomes rapidly too big
  for large $\tb$, as shown in Fig. \ref{f2}.
  Note the change in sign of $\mu$ for the positive (negative) interference
  between LO and NLO in the figure.

The scale dependence in NLO is still large, as shown by the width of
the bands in
Fig. \ref{f2}, which may be related to the incomplete NLO calculations.
Due to the large scale uncertainty one expects a good fit 
for practically all values
of $\tb$ and $\mu<0$, if the scale is left free in the fit within
the limits of 0.5 and 2 $m_b$.
This is shown in Fig. \ref{f6}, where only a very small corner is excluded
by \besg in contrast to the previous LO calculations\cite{PL}.
%The low $\tb$ scenario from Ref. \cite{PL} is shown for comparison.
%Previously it was constrained mostly by the relic density,
%although now the whole low \tb scenario is excluded by the present
%limits on the Higgs mass from LEP, as will be discussed in the next
%section.

In Ref. \cite{ciu} only the effect of the ligtest stop to the NLO
contributions was considered and no flavour mixing between
the three generations was taken into account.
The latter was found to be small in the CMSSM.
The effect of the missing contributions of the heavier stop has been
studied. The general picture does not
change by including  heavier stop terms analogous to the light stop terms,
although complete calculations including diagrams where both stops
contribute simultaneously have not yet been calculated.
Given the uncertainties from the incomplete calculations, we will exclude
hereafter  the \besg constraint from the fit and study the Higgs mass
prediction in the CMSSM.

\section{Higgs mass predictions}

In Supersymmetry the couplings in the Higgs potential
are the gauge couplings. The absence of arbitrary couplings
together with well defined radiative corrections to the masses
results in clear predictions for the lightest Higgs mass
and electroweak symmetry breaking (EWSB).

In the Born approximation one expects the lightest Higgs to have a mass $m_h$ below the $Z^0$ mass. However, loop corrections, especially from top and stop quarks, can increase $m_h$ considerably. 
The Higgs mass depends mainly on the following parameters: the top mass,
the squark masses, the mixing in the stop sector,
the pseudoscalar Higgs mass and $\tan\beta$. 
As will be shown below, the maximum Higgs mass is obtained for 
large $\tan\beta$, for a maximum value of the top and squark masses and 
a minimum value of the stop mixing.
The Higgs mass calculations were carried out following the results
obtained by 
Carena, Quir\'os and Wagner\cite{carenawagner} in a renormalization 
group improved effective potential approach,  including the 
dominant two-loop contributions from gluons and gluinos.
%The gluino contributions were taken from the FeynHiggs
%calculutions\cite{feynhiggs}.

Note that in the CMSSM the Higgs mixing parameter $\mu$ is determined 
by the requirement of EWSB, which yields large values for $\mu$\cite{rev}. 
Given that the pseudoscalar Higgs mass increases rapidly with $\mu$, this mass is always much larger than the lightest Higgs mass and thus decouples. 
We found that this decoupling is effective
for all regions of the CMSSM parameter space, i.e. the lightest Higgs 
has the couplings of the SM Higgs within a few percent. 
Consequently the experimental  limits on the SM Higgs can be taken.

The lightest Higgs boson mass $m_h$ is
shown as function of $\tan\beta$  in
Fig.~\ref{mhiggstanb}. The shaded band corresponds
to the uncertainty from the stop mass and stop mixing for $m_t=175$ GeV.
The upper and lower lines correspond to $m_t$=170 and 180 GeV, respectively. 

%The parameters used for the calculation of the
%upper limit were: $m_t=180$ GeV, $A_0=-3m_0$ and $m_0=m_{1/2}=1000$ GeV.
One observes that for a SM Higgs limit of 107.9 GeV \cite{newhiggs} all
values of \tb below 3.3 are excluded in the CMSSM. 
%The lowest line of the
%same figure gives the minimal values  of $m_h$. For  high
%$\tan\beta$  the values of $m_h$ range from 105 GeV
%to 125 GeV. 
%There is at present no preference for any of the values
%in this range, but  it can be seen, that the 95\% C.L. lower limit
%on the Higgs mass\cite{newhiggs} excludes $\tb<3.3$. 
% we calculate the standard deviation around the central
%value, assuming a flat prior in the interval and find:
%$$\rm m_h=115\pm5.8~\rm GeV.$$ 

In order to understand better the Higgs mass uncertainties,
 the relevant parameters were varied one by one.
%The largest uncertainty on the light Higgs mass 
%originates from the stop masses, as shown in Fig. \ref{hi35}
%for $\tb=35$, $A_0=0$ and $m_t=175$ GeV.
The Higgs mass varies between 110 and 120 GeV, if $m_0$ and $m_{1/2}$
are varied between 200 and 1000 GeV, which implies stop masses
varying between 400 and 2000 GeV, as shown inf Fig. \ref{hi35}.
Since at present there is no preference for any of the values
between 110 and 120 GeV, the variance for a flat probability 
distribution is 10/$\sqrt{12}$=3 GeV, which we take as an error estimate.

%The remaining uncertainty on the Higgs mass originates from
%the mixing in the stop sector when one leaves $A_0$ a free parameter. 
%The mixing is determined by the off-diagonal
%element in the mass matrix $X_t=A_t-\mu/\tb$.
The dependence of the Higgs mass on $A_0$ is shown in Fig. \ref{wing} for the
high $\tan\beta$ scenario. The influence on the Higgs mass
is quite small in the CMSSM, since the low energy value $A_t$ tends
to a fixed point, so that the stop mixing parameter $X_t=A_t-\mu/\tb$
is not strongly dependent on $A_0$. Furthermore, the $\mu$ term is
not important at large $\tb$.
If we vary $A_0$ between $\pm3m_0$, the  
 the error from the stop mixing in the Higgs boson mass 
is estimated to be $\pm 1.5$ GeV. The values of $m_0=m_{1/2}=370$ GeV 
yield the central value of $m_h=115$ GeV.

The dependence on $m_t$ is shown in Fig. \ref{mtopmh} for $A_0=0$ and 
intermediate values of $m_0$ and $m_{1/2}$ for two values of $\tb$ 
(corresponding to the minimum $\chi^2$ values in Fig. \ref{tbmtop}). 
The uncertainty from the 
top mass at large $\tb$ is  $\pm$ 5 GeV, given the
uncertainty on the top mass of 5.2  GeV.

The uncertainties from the higher order calculations (HO) is estimated to be
2 GeV from a comparison of the full diagrammatic method \cite{feynhiggs}
and the effective potential approach\cite{carenawagner}, 
so combining all the uncertainties discussed before we find for
the  Higgs mass in the CMSSM
$$m_h=115\pm3~ (stop mass)~\pm1.5~(stop mixing)~\pm2~(theory)~\pm5~(top
mass)~\rm GeV.$$
where the errors  are the estimated standard deviations around the central
value.
As can be seen from Fig.~\ref{mhiggstanb} this  central value 
is valid for all $\tan\beta > 20$ and decreases for lower $\tan\beta$.
\section{Higgs  and Chargino mass versus LSP mass}
In the CMSSM considered here all masses are related, since
they have a common mass at the GUT scale. The low energy values are
completely determined by the renormalization group equations\cite{rev}.
If R-parity is conserved, the lightest supersymmetric particle (LSP)
will be stable, which is usually the lightest neutralino.
This LSP has all the
properties of a WIMP (Weakly Interacting Massive Particle)
 required for the cold dark matter in our universe{cite{rev}.

The contours of the Higgs mass, chargino mass  and LSP mass are shown
in Fig. \ref{lsp}. The LSP is practically independent of $m_0$ and
is given by $\approx 0.4 m_{1/2}$\cite{rev}, so an LSP of $52^{+10}_{-8}$
which could be the interpretation of the annual modulation signature
observed by the DAMA Collaboration\cite{dama},
corresponds to $m_{1/2}\approx 130^{25}_{20}$ GeV.
As can be seen from Fig. \ref{lsp}, the central value of 130 GeV is excluded by  a Higgs mass below 107.9 GeV
for $m_0$ below 750 GeV and for a chargino mass below 100 GeV
for the higher $m_0$  values.

For this year of LEP running one hopes to be sensitive to Higgs masses
up to 114 GeV, i.e. LSP masses up to 120 GeV,
 which clearly will be able to answer the question,
if the observation by the DAMA collaboration can be interpreted
as a CMSSM LSP.

\section{Conclusions}

The results can be summarized as follows:
\begin{itemize}
\item
The NLO \besg contributions change the preferred sign of the Higgs mixing
parameter $\mu$ for the large $\tb$ scenario of the CMSSM. Since this
sign agrees now with the sign preferred from $b-\tau$ unification,
the allowed parameter becomes much larger. Given the larger uncertainties
from the still incomplete calculations, we conclude that at present 
no constraints from \besg can be derived.
\item
The low $\tb$ scenario ($\tb<3.3$) of the CMSSM is excluded by the
 lower limit on the Higgs mass of  107.9 GeV\cite{newhiggs}.
\item
For the high $\tb$ scenario the  Higgs mass is found to be in
the range from 110 to 120 GeV for $m_t=175$ GeV.
 The errors around the central
value of 115 GeV are found to be:
$m_h=115\pm3~ (stop mass)~\pm1.5~(stop mixing)~\pm2~(theory)~\pm5~(top
mass)~\rm GeV.$
This prediction is independent of $\tb$ for $\tb>20$
and decreases for lower $\tb$.
\item
The interpretation of the annual modulation signature by the DAMA Collaboration
as the lightest neutralino in the CMSSM is at the edge of the parameter space
allowed by the present Higgs and chargino limits.
Future running at LEP will be able to settle the question, if
an LSP of $52^{+10}_{8}$ GeV can be accomodated in the CMSSM.
\end{itemize}

\section*{Acknowledgements}

We thank P. Gambino, G.F. Giudice and M. Misiak for helpful 
discussions on the NLO \besg rates.

\begin{figure}[htb]
\epsfig{file=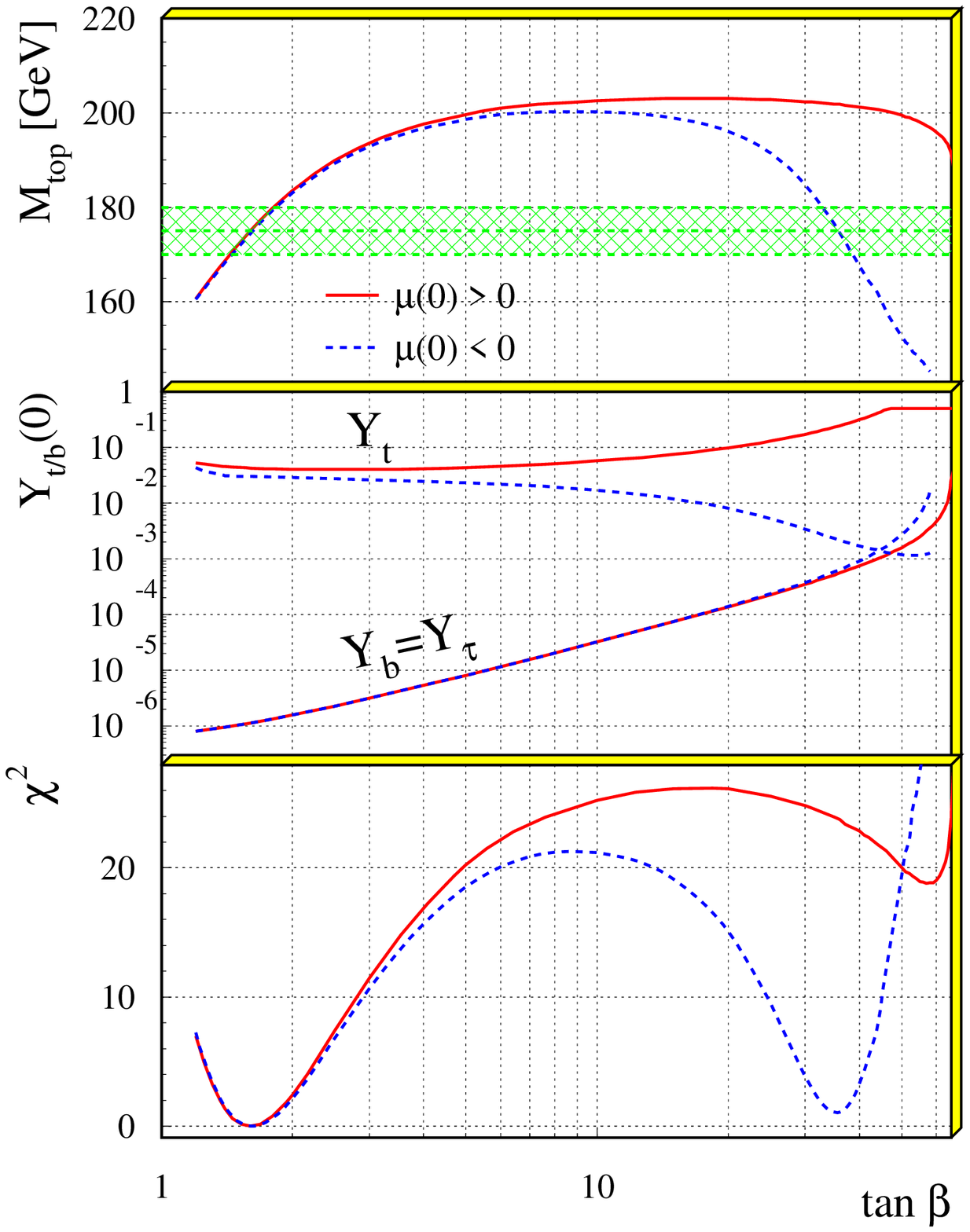,width=\textwidth}
\caption[]{\label{tbmtop} 
The upper part shows the top quark mass as function of tan $\beta$ for $m_0$ = 600 GeV,$m_{1/2}$ = 400 GeV. The middle part shows the corresponding values of the Yukawa couplings at the GUT scale and the lower part the $\chi^2$ values. As can be seen the value of tan $\beta$ is restricted to be in the following ranges 1$ < \tan\beta < $2 \  or 30$ < \tan\beta < $40 for $\mu < 0$.}
\end{figure}
\begin{figure}[htp]
\epsfig{file= 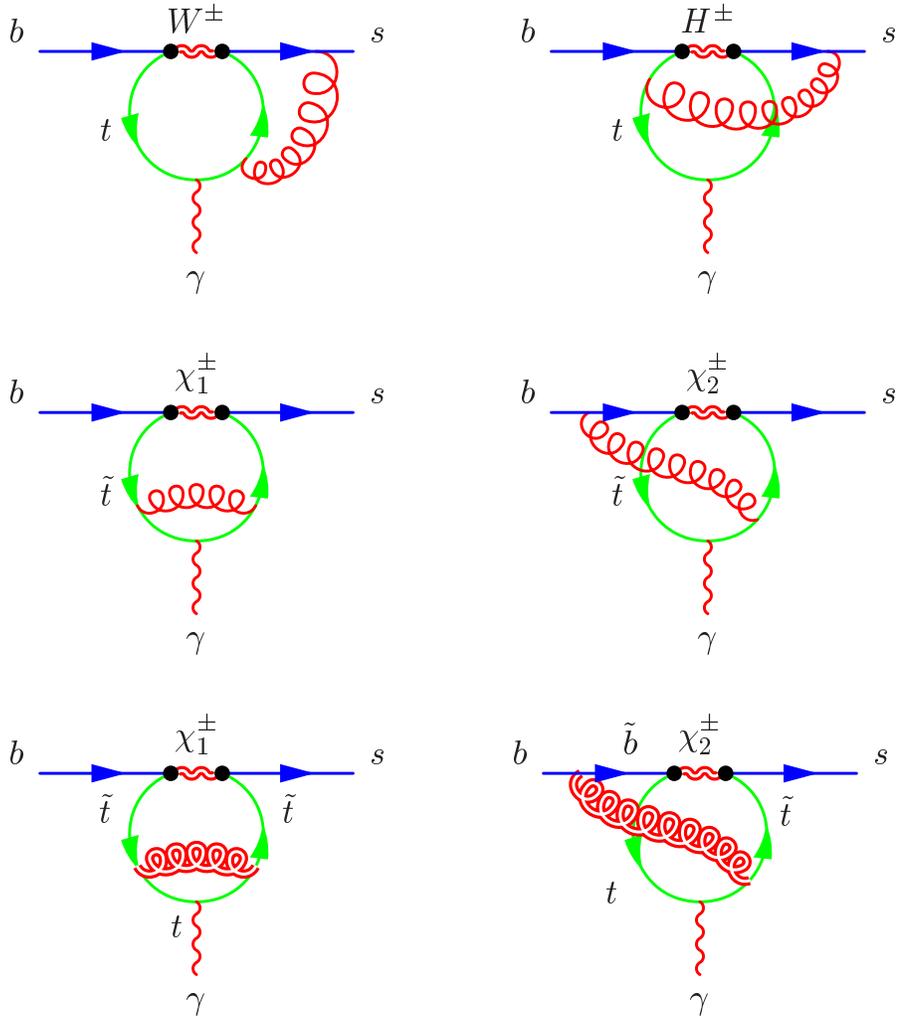,width=\textwidth}
\caption[]{\label{f1} 
Some electroweak loop diagrams for the \besg \ transition in NLO.}
\end{figure}
%
%--------------- $b\to s\gamma$ - amplitudes,mu<0
%
\begin{figure}[t]
\epsfig{file=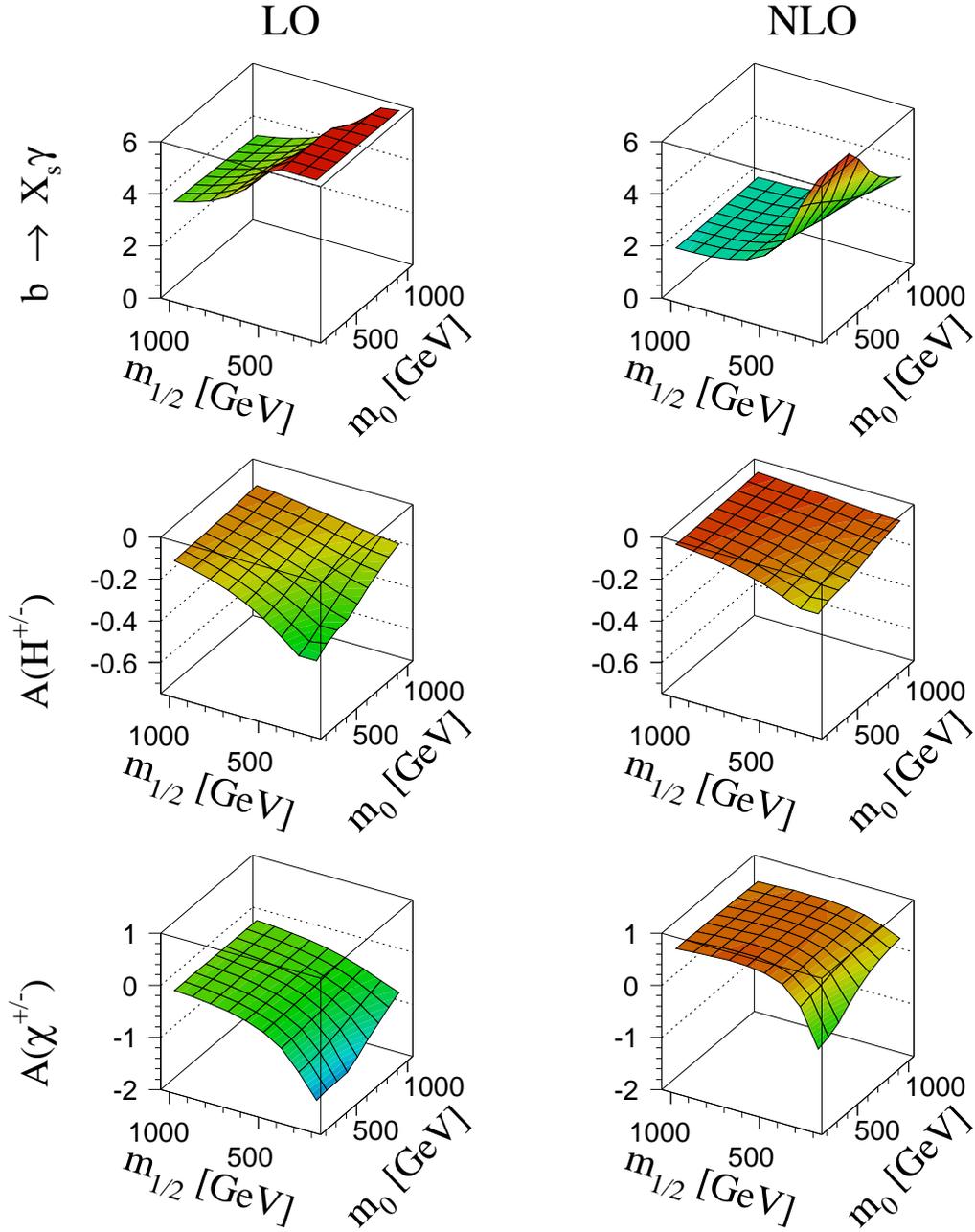,width=\textwidth}
\caption[]{\label{f3} 
The decay rate (in units of $10^{-4}$) and selected amplitudes
(in units of $10^{-2}$) of the \besg \  decay for 
negative $\mu$ and $\tan\beta = 35$.
These amplitudes should  be  compared with the
SM amplitude of $-0.56\cdot 10^{-2}$. Note the sign change in the $\chi^{\pm}$ amplitude.}
\end{figure}

\begin{figure}[t]
\epsfig{file=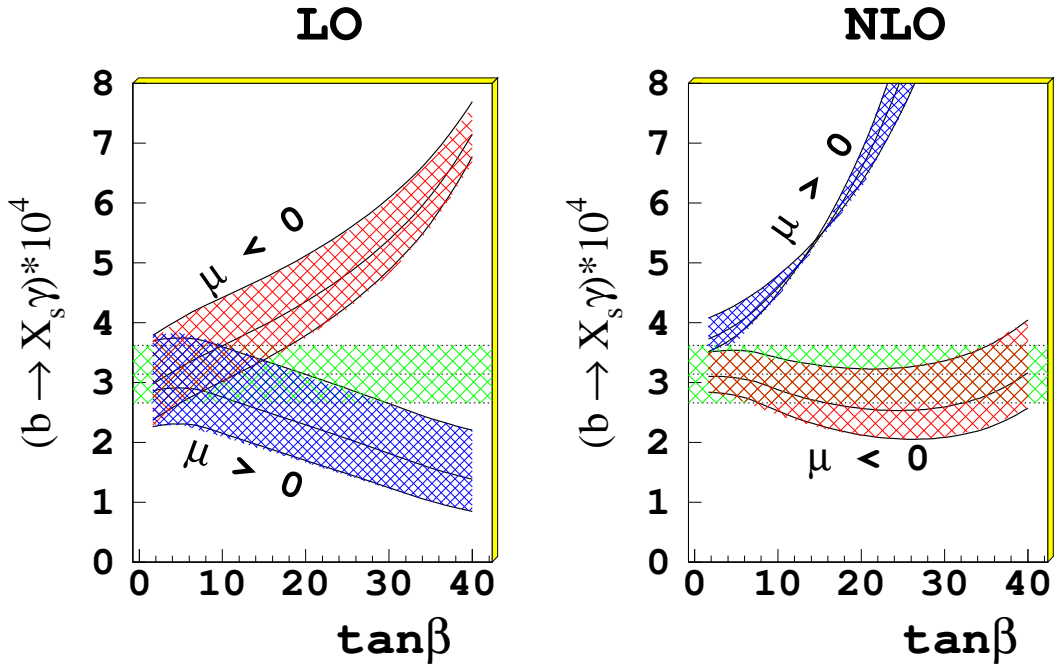,width=\textwidth}
\caption[]{\label{f2} 
The dependence of the \besg \ rate on $\tan\beta$ for LO (l.h.s.)
and NLO (r.h.s.) for m$_0$ = 600 GeV and m$_{1/2}$ = 400 GeV. 
A fit was made for each value of $\tan\beta$ and $\mu$. The
curved bands show  the theoretical prediction; its width is
determined by the renormalization scale uncertainty, if
this is varied between 0.5m$_b$ and 2m$_b$.
The horizontal band shows the
experimental value  $\pm 1\sigma$.
}
\end{figure}
\begin{figure}[htp]
\begin{center}
\epsfig{file=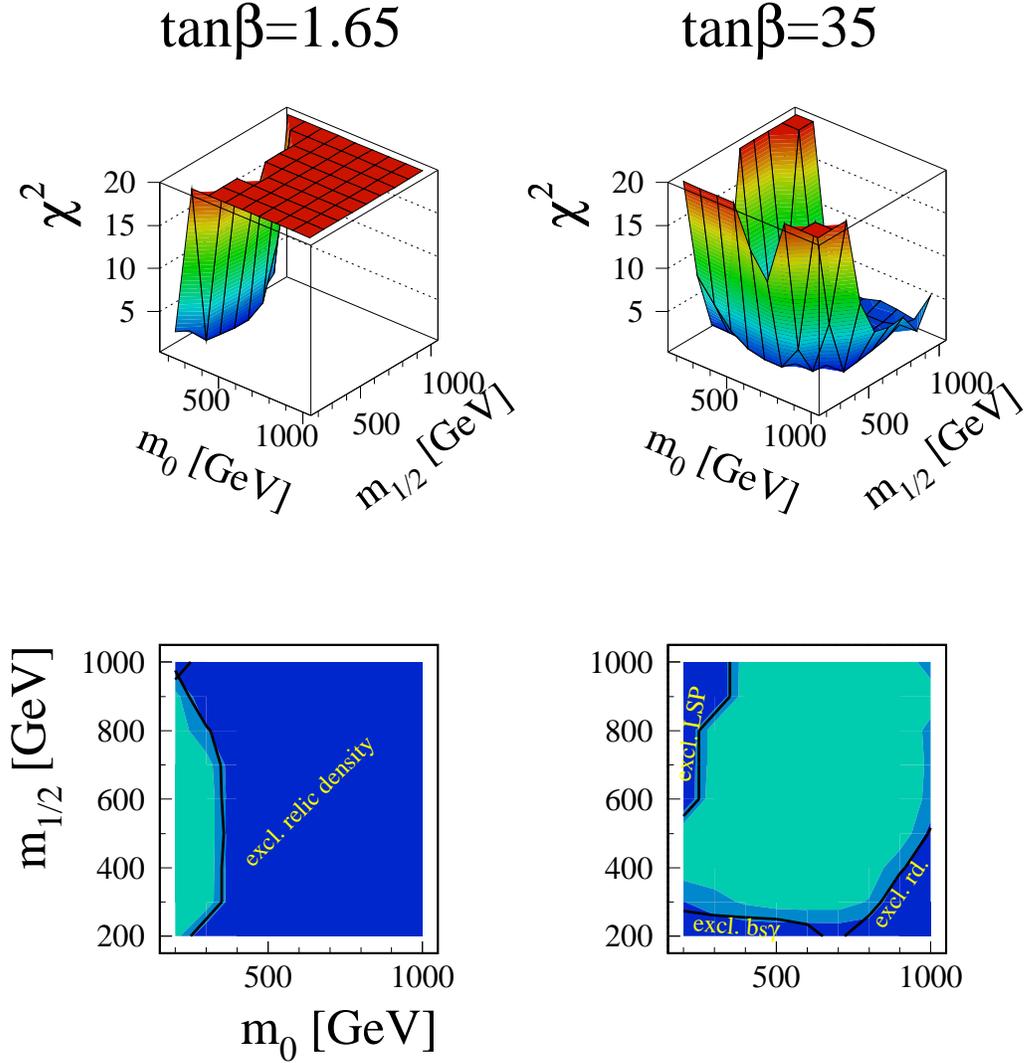,width=\textwidth}  
\caption[]{\label{f6}
The upper row shows the $\chi^2$ distribution in the $m_0-m_{1/2}$ plane
for $\tan\beta = 1.65$ and $\tan\beta = 35$.
The projections are shown in the second row.
The different shades in the projections
indicate steps of $\Delta\chi^2=4$. The contour lines show areas
excluded by the particular constraints used in the analysis:
in the LSP area the Lightest Supersymmetric Particle is charged
(usually the stau),
which is not allowed if the LSP is stable,
 in the relic density (rd) area the density of the universe is above
 the critical density and in the $bs\gamma$ area the \besg rate is too high.
In the $\tan\beta = 35$ plots the $\mu_b$-scale was allowed to vary
between 0.5m$_b$ and 2m$_b$ for the \besg rate.
}
\end{center}
\end{figure}
\begin{figure}[htb]
\begin{centering}
\epsfig{file=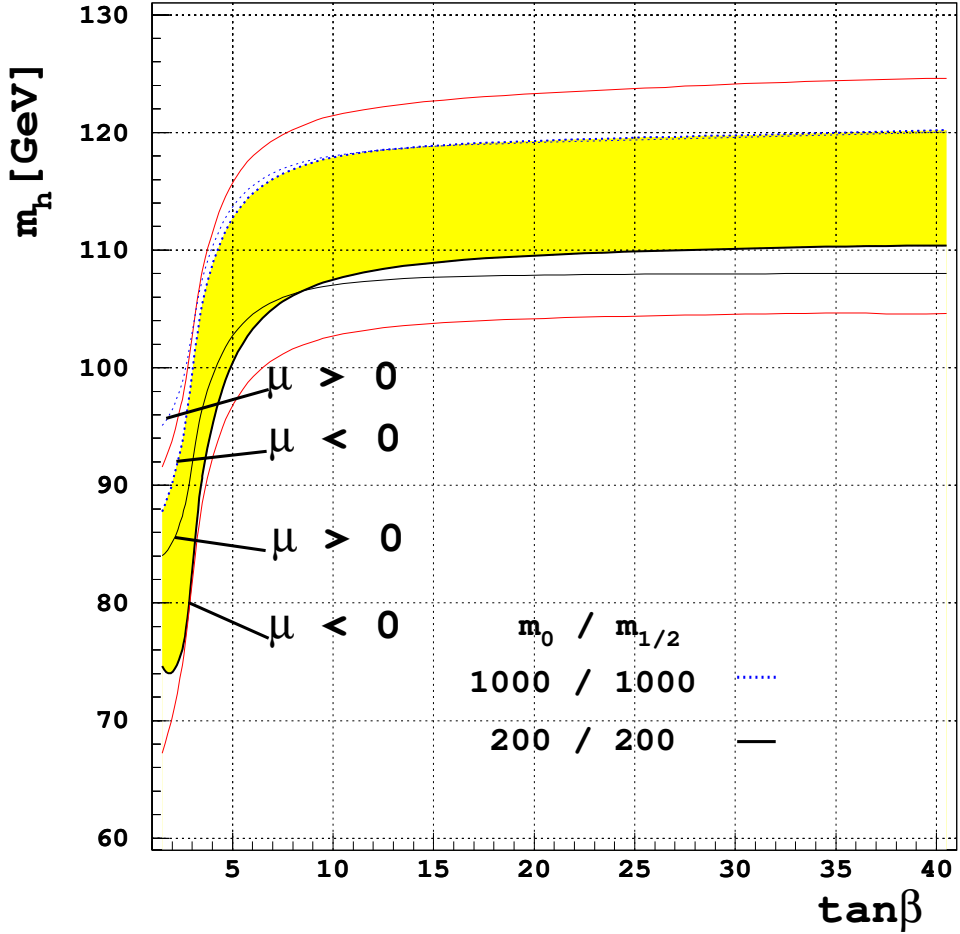,width=.95\textwidth}  
\caption[Dependence of the mass of the lighter {$\cal CP$}-even Higgs-boson on $\tan\beta$]{\label{mhiggstanb}
%The figure shows for both signs of the Higgs mixing parameter $\mu$ 
The dependence of the mass of the lighter {$\cal CP$}-even Higgs boson on
$\tan\beta$, as calculated by the effective potential
approach\cite{carenawagner}.
The shaded band shows the variation of $m_0=m_{1/2}$ between 200 and 1000 GeV 
for $\mu<0$, $m_t=175$ GeV, and $A_0=0$. Note the small dependence
on the sign of $\mu$ for large $\tb$, as expected from the
suppression of $\mu$ by \tb in the stop mixing.
The maximum (minimum) Higgs boson mass value, shown by the upper (lower) line
are obtained for $A_0=-3m_0$, $m_t=180$ GeV, $m_0=m_{1/2}=1000$ GeV
($A_0=3m_0$, $m_t=170$ GeV, $m_0=m_{1/2}=200$ GeV).
As can be seen the curves show an asymptotic behaviour for large values
of $\tan\beta$.
} \end{centering} \end{figure}
\begin{figure}[t]
\epsfig{file=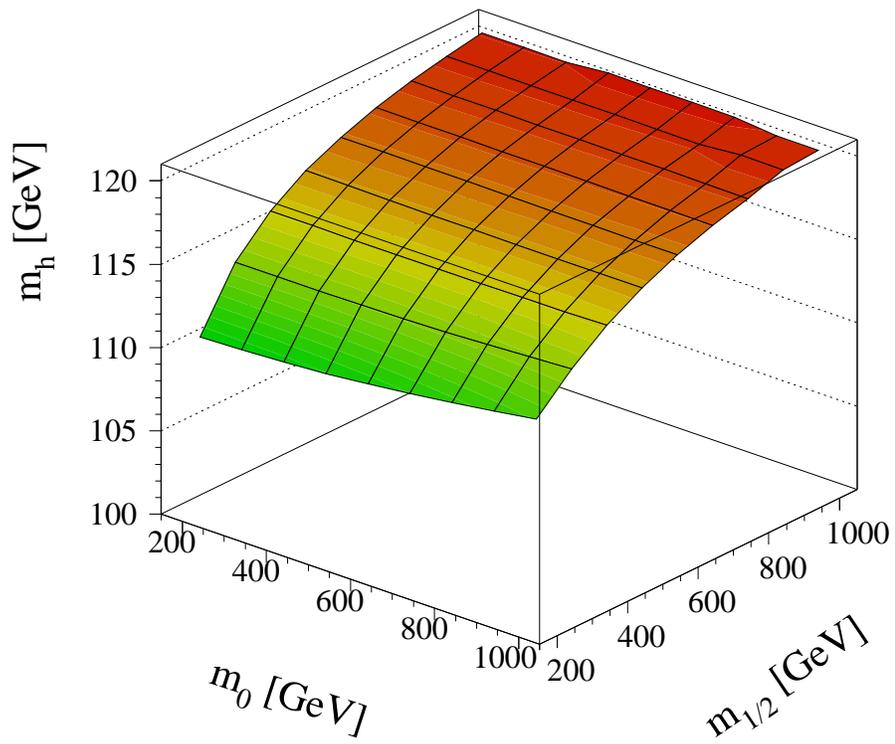,width=\textwidth}
\caption[]{\label{hi35}
The Higgs boson mass as function as function of $m_0$ and $m_{1/2}$,
as calculated by the
effective potential approach\cite{carenawagner}. 
 }
\end{figure}
\begin{figure}[htb]
\begin{centering}
\epsfig{file=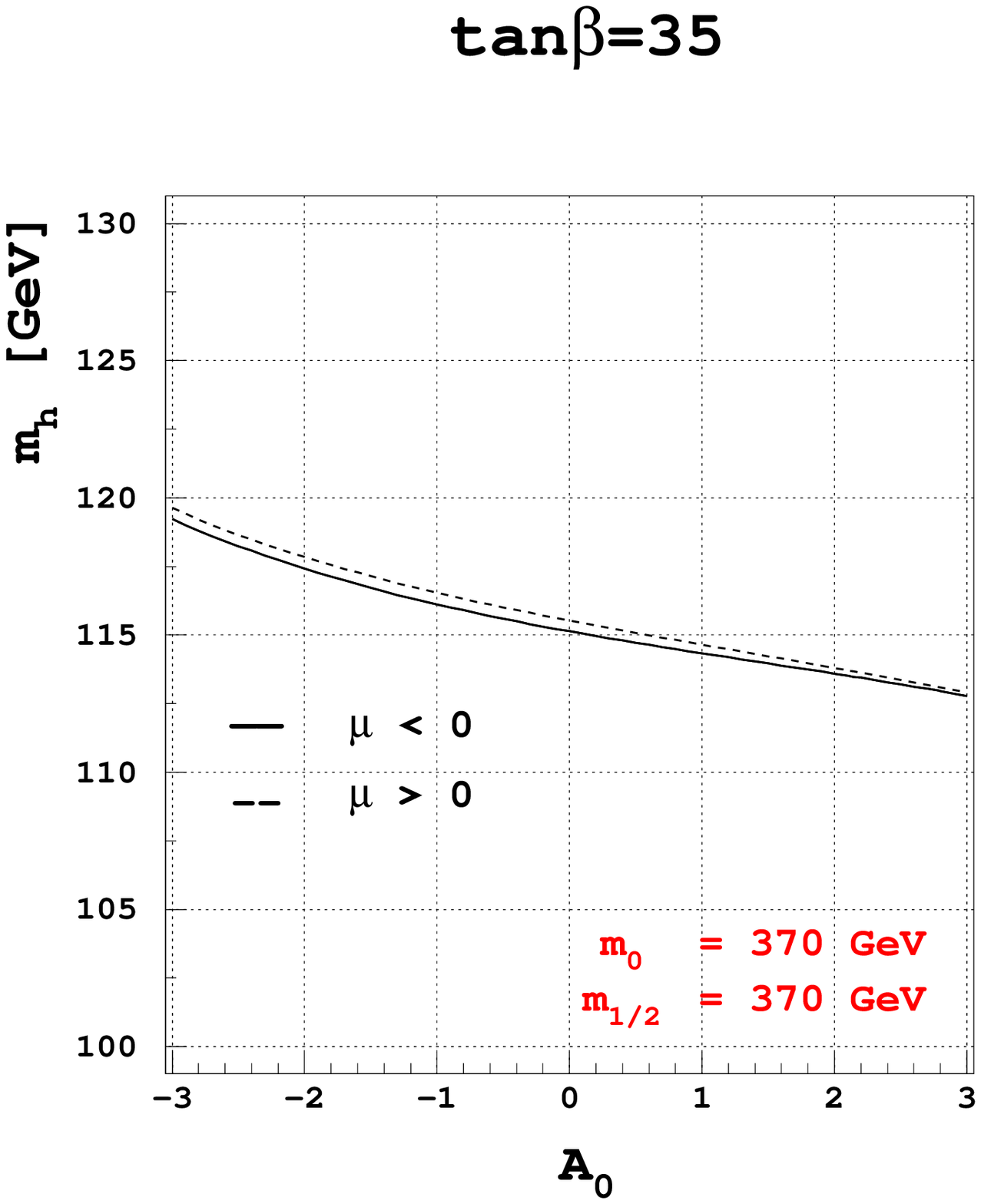,width=\textwidth}  
\caption[Dependence of the Higgs mass on $A_0$]{\label{wing}
Dependence of the Higgs mass on the trilinear coupling $A_0$
at the GUT-scale in units of $m_0$. 
The low-energy value $A_t$ tends to a fixed point,
thus reducing the  influence of $A_0$ on the mixing in the stop sector.
}
\end{centering}
\end{figure}
\begin{figure}[htb]
\begin{centering}
\epsfig{file=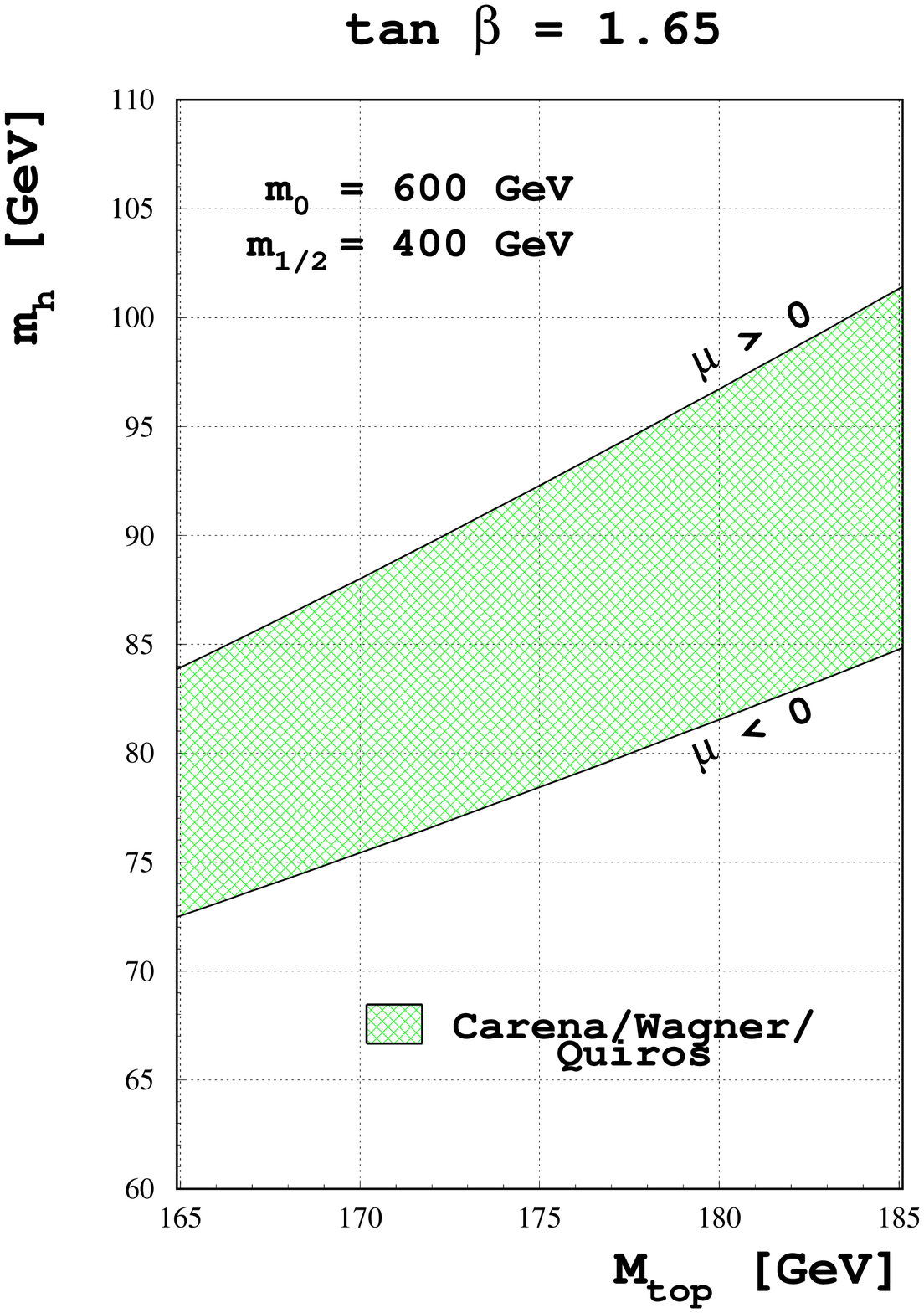,width=.49\textwidth}  
\epsfig{file=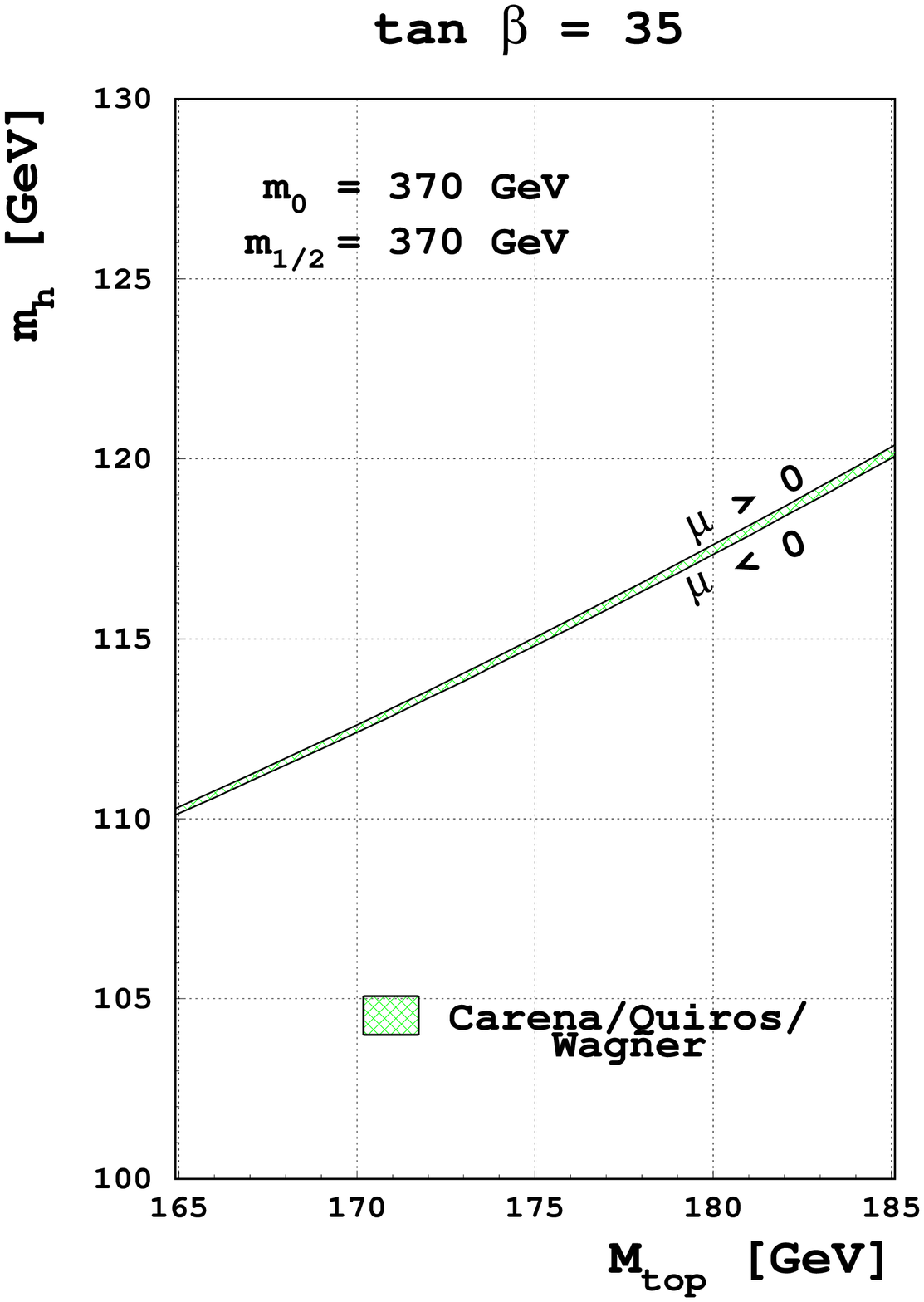,width=.49\textwidth}  
\caption[Dependence of the Higgs mass on $M_{\rm top}$]{\label{mtopmh}
The top mass dependence of the Higgs mass  in the low and high $\tb$ scenario.
Note the reduced dependence on the sign of $\mu$ for large \tb, as
expected from the stop mixing parameter $X_t=A_t-\mu/\tb$.
}\end{centering}
\end{figure}
\begin{figure}[htb]
\begin{centering}
\epsfig{file=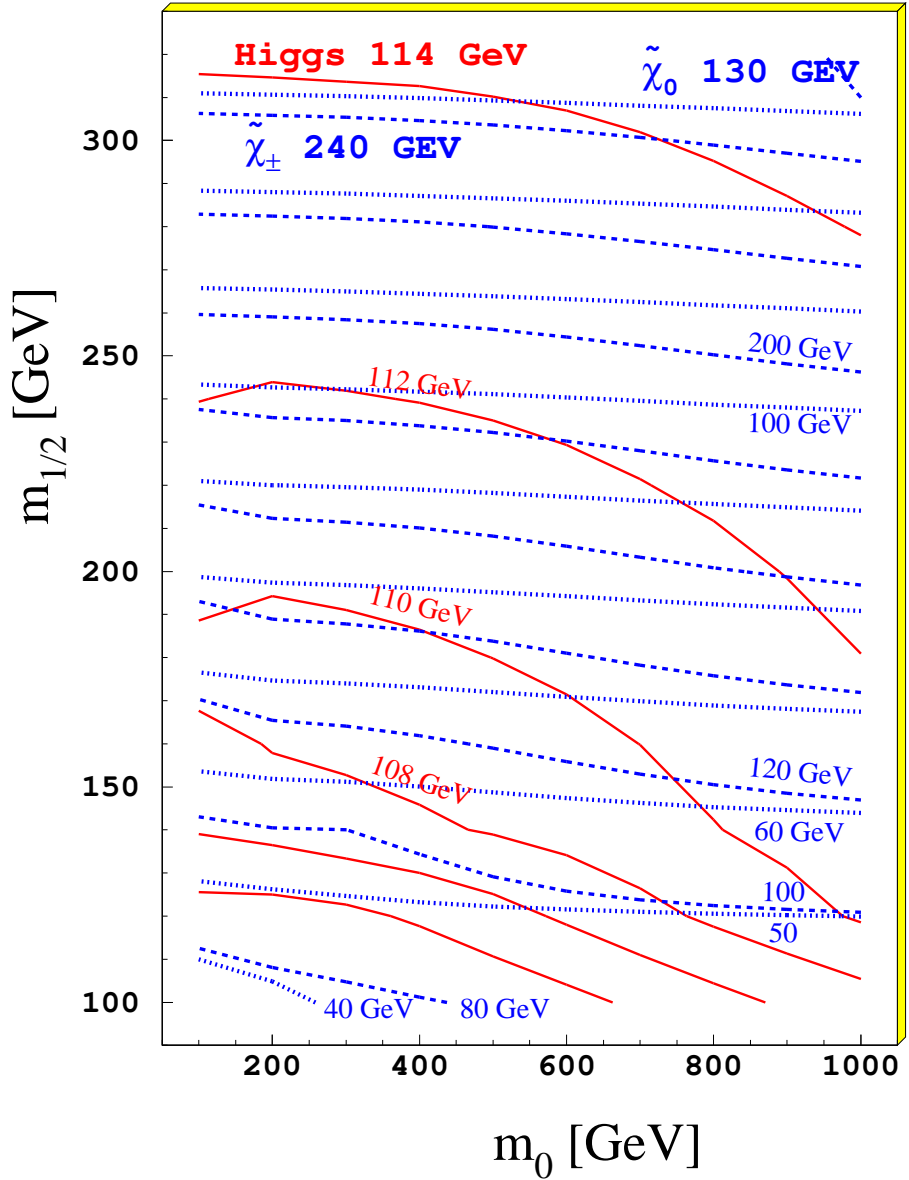,width=\textwidth}
\caption[higgs vs LSP mass]{\label{lsp}
Contours of the  Higgs mass (solid lines), lightest chargino mass (dashed) and
LSP mass (fine dashed)  in the $m_0,m_{1/2}$ plane in steps of 2,10 and 20 GeV.
The region for a Higgs mass below 107.9 GeV and chargino mass below 100 GeV
is excluded. It can be seen that an 
LSP mass of 52 GeV is excluded too, so  the DAMA result interpreted
as an LSP with a mass of $52^{+10}_{-8}$\cite{dama} is at
the edge of the allowed parameter space.
 This plot is 
for large $\tb$; for smaller $\tb$ the exluded region rapidly increases
because of the Higgs mass limit.
}
\end{centering}
\end{figure}

\end{document}